\documentclass{appolb}
\usepackage{graphicx}

\begin{document}
\title{EVENT BY EVENT FLUCTUATIONS OF THE SOURCE SHAPE:
IMPLICATIONS FOR THE LEVY SHAPE,\\ AND EVENT SHAPE SORTING%
\thanks{Presented at the XIII Workshop on Particle Correlations and Femtoscopy, Krak\'ow, 22-26 May 2018}%
}
\author{Boris Tom\'a\v{s}ik
\address{Univerzita Mateja Bela, Tajovsk\'eho 40, 97401 Bansk\'a Bystrica, Slovakia, \\and\\
FNSPE, \v{C}esk\'e vysok\'e u\v{c}en\'i technick\'e v Praze, B\v{r}ehov\'a 7, 11519 Praha 1, Czechia}%
\\
\vspace{1em}
Jakub Cimerman
\address{FNSPE, \v{C}esk\'e vysok\'e u\v{c}en\'i technick\'e v Praze, B\v{r}ehov\'a 7, 11519 Praha 1, Czechia}
}
\maketitle
\begin{abstract}
In the first part of this contribution we show that the Levy stable shape of the correlation function can be caused 
by averaging of the measured correlation functions over a large number of events. In the second part it is demonstrated
how a sample of events sorted by Event Shape Sorting technique exhibits different azimuthal dependence 
of correlation radii in each event class. 
\end{abstract}
  
  
\section{Motivation}
It is widely understood that in the experiments with relativistic nuclear collisions each event evolves differently 
from the others. This is true even if carefully selected centrality classes are studied.  Event by event fluctuations
of the hadron distributions and its anisotropies, parametrised through the coefficients $v_n$, clearly demonstrate this. 

In spite of that, in femtoscopic studies source sizes averaged over a very large number of events are always extracted
and no attention is paid to their fluctuations. The impact of the fluctuations on the resulting event-averaged 
correlation function have recently been investigated in \cite{Plumberg:2015mxa}. 

We want to make two points in this short contribution: i) The averaging over a large number of different sources 
influences the shape of the correlation function and may result in a shape given by Levy-stable distribution. ii) The 
fluctuations of sizes and the space-time characteristics of the sources can be accessed through the Event Shape 
Sorting technique \cite{Kopecna:2015fwa}.


\section{Levy-stable distribution from averaging over events}

The measured correlation function is often non-Gaussian and fitted with Levy stable distribution
\begin{equation}
C(q) - 1 = \lambda \exp \left ( - | Rq |^\alpha \right )
\label{e:LSd}
\end{equation}
where $q$ is momentum difference of the pair and $R$, $\alpha$, and $\lambda$ are fit parameters. It has been pointed out already in 
\cite{Bialas:1990gt} that such a shape may result from summing up production from sources with certain distribution 
of sizes. We recall that even on a single event we have effectively many sources with different sizes, since each pair of bosons 
measures the size of the homogeneity region specified by its momentum. The averaging is on the level of the correlation 
function. 

\begin{figure}
\centerline{\includegraphics[width=0.45\textwidth]{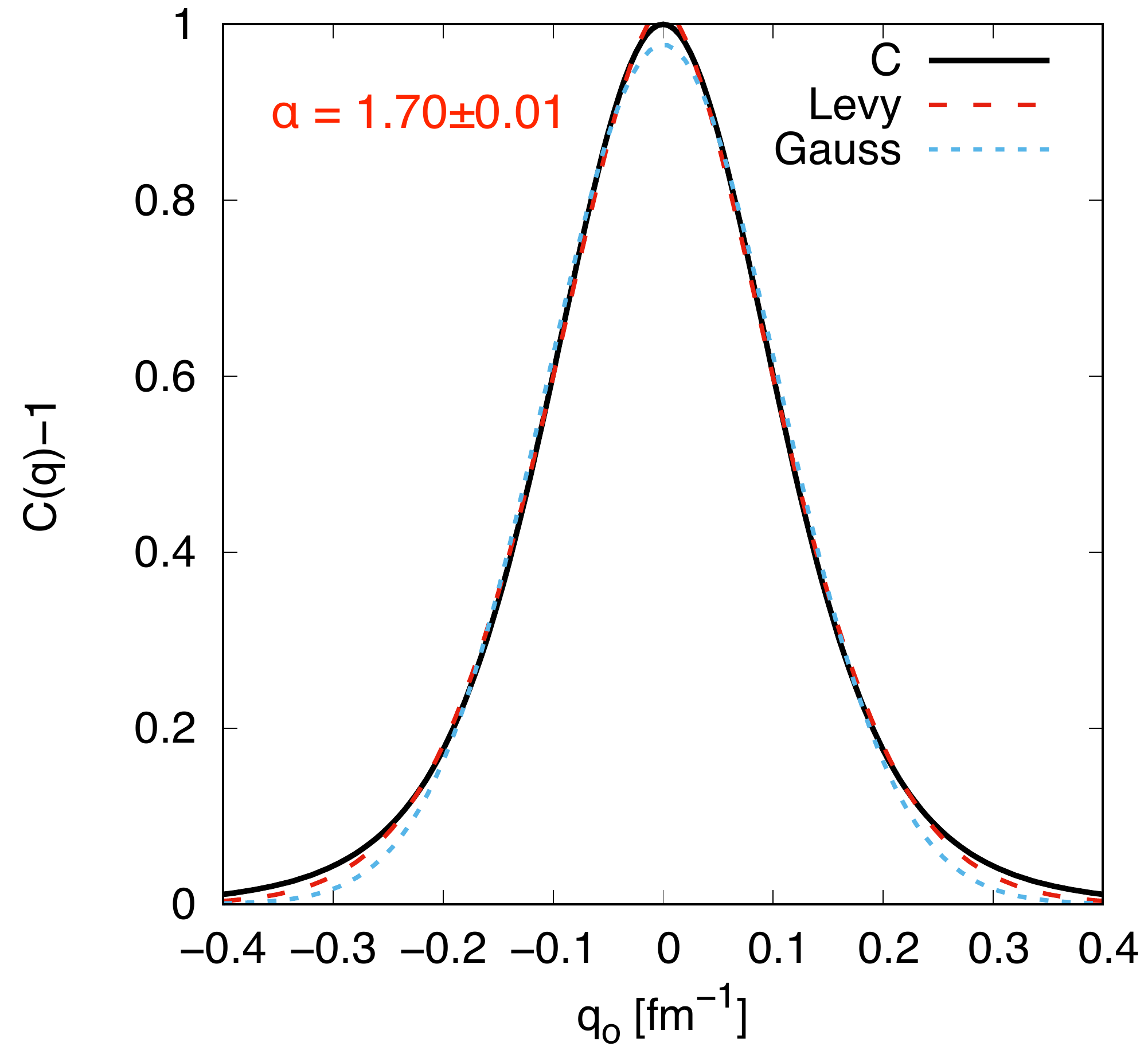}%
\includegraphics[width=0.45\textwidth]{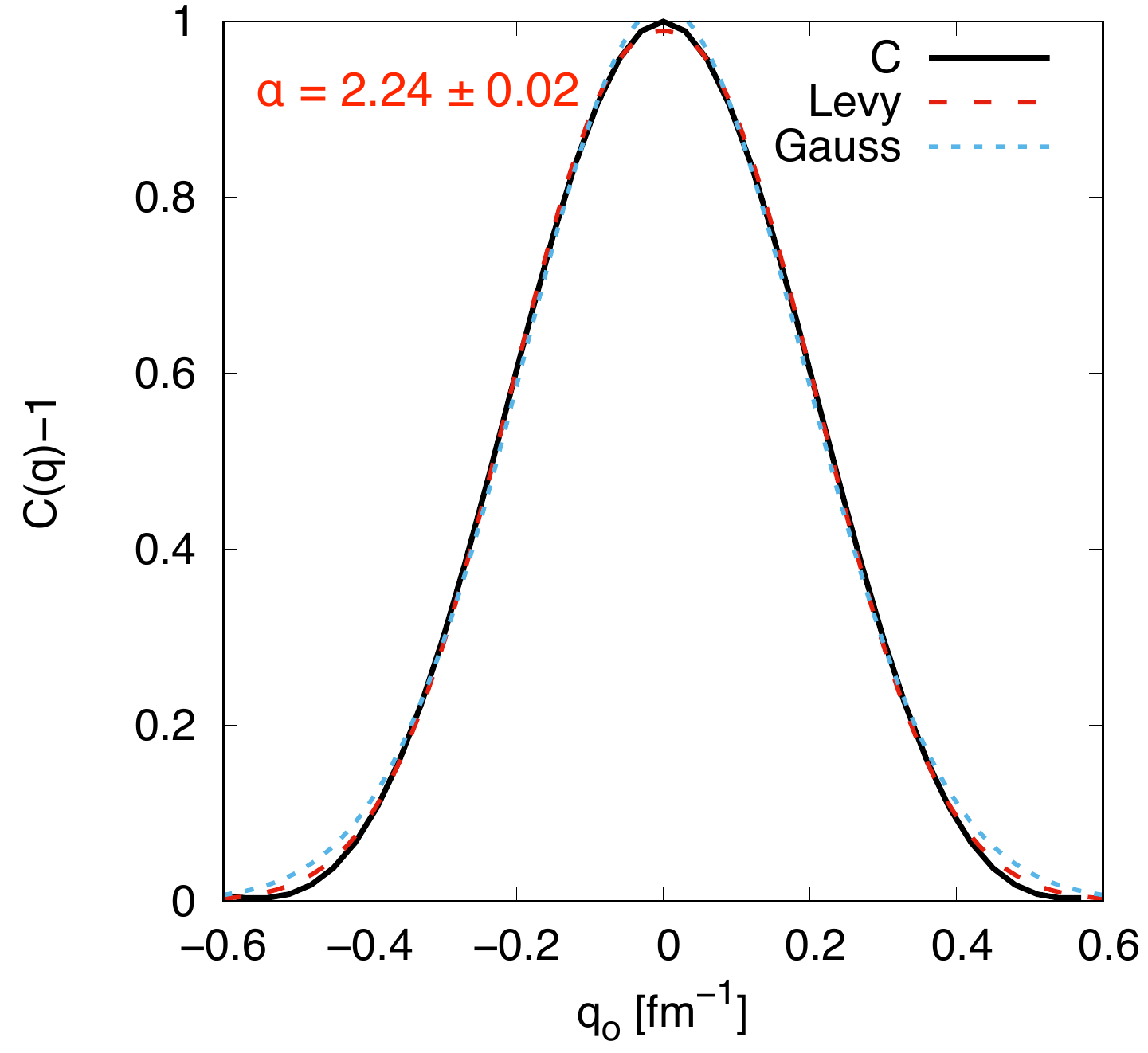}}
\caption{The shapes of the correlation function as resulting from averaging over source sizes and orientations. Black solid line is
the calculated correlation function, red dashed line is the best fit with distribution (\ref{e:LSd}) with the parameter $\alpha$ indicated 
in the figure, and blue dotted line is the best fit with Gaussian distribution. 
\label{f:Caver}
}
\end{figure}

We demonstrate this with two examples. First, let us assume a simple toy-model source, which is given in transverse plane  by
two-dimensional Gaussian profile with sizes $R_1$ and $R_2$. Its axes are rotated with respect to the Cartesian $x$, $y$ system 
by the angle $\theta_2$. The correlation function from such a source is calculated the usual way 
\[
C(q) -1 = \frac{\left | \int d^4x\, S(x) e^{iqx}\right |^2}{\left ( \int d^4x\, S(x) \right )^2}\,  .
\]
Such a correlation function will be Gaussian. To represent the averaging over sources which have different sizes and are differently 
oriented, we then integrate over correlation functions which resulted from all the different $R$'s. They have been distributed 
like the two sizes of the overlapping region according to the optical Glauber model, and the rotation angle $\theta_2$ is 
distributed uniformly. The resulting correlation function is no longer Gaussian. In Fig.~\ref{f:Caver} (left) we show its cut along 
the outward (or $x$) direction. 
We see that just from such a simple averaging we obtain a correlation function which is best reproduced by the index parameter
about $\alpha = 1.70$.

In our second example we use a source given by the blast-wave model. This version of the blast-wave model is extended to describe 
transverse anisotropies to an arbitrary order \cite{Cimerman:2017lmm}, although here we will only go up to the second order.
The important feature of the model is that it incorporates the anisotropies of the transverse size of the fireball $R(\varphi)$ 
and also the anisotropies of the transverse flow, with the help of the flow rapidity in the transverse direction $\rho(r,\varphi)$
\begin{eqnarray}
R(\varphi) & = & R_0 \left ( 1 - \sum_{n=2}^\infty a_n \cos\left ( n (\varphi - \varphi_n)\right ) \right )\,  , 
\label{e:Rani}
\\
\rho(r,\varphi) & = & \frac{r}{R(\varphi)} \rho_0 \left (
1 + \sum_{n=2}^\infty 2\rho_n \cos\left ( n (\varphi - \varphi_n)\right ) 
\right )\,  .
\label{e:flowani}
\end{eqnarray}
We have calculated the dependence of the correlation function on $q_o$ and $q_s$ from this model with $T=120$~MeV, $R_0 = 7$~fm, $\rho_0 = 0.8$ 
and the anisotropy parameters are $a_2 = \rho_2 = 0.2$. The cut through the correlation function along the $q_o$ axis is shown in Fig.~\ref{f:Caver} 
(right).  While the best fit function belongs to the family (\ref{e:LSd}), it is no longer a Levy stable
distribution, which is limited to $\alpha < 2$.

We summarise this section with the conclusion, that the averaging over many different sources in the determination of the measurable 
correlation function may profoundly influence the actual shape of the correlation function.


\section{Event Shape Sorting and femtoscopy}

The recently proposed method of Event Shape Sorting (ESS) \cite{Kopecna:2015fwa} may actually help to select events more selectively 
and avoid some of the averaging. The algorithm sorts the events in such a way, that events with similar azimuthal distribution 
of hadrons end up close together. They may be assumed to have undergone similar fireball evolution. Except for femtoscopic studies, such 
event selection may be interesting also for studies of jet quenching, since it may select events according to the actual azimuthal shape of the fireball
which is relevant for the path length of the leading parton.


\subsection{The difference to Event Shape Engineering}

\begin{figure}
\centerline{
\begin{minipage}[b]{0.16\textwidth} \
\end{minipage}
\begin{minipage}[b]{0.18\textwidth} 
\begin{footnotesize}
$$
v_2 = 0.04
$$ \vspace{-0.07\textheight}$$
v_3 = 0.04
$$
\end{footnotesize}
\end{minipage}
\begin{minipage}[b]{0.18\textwidth} 
\begin{footnotesize}
$$
v_2  =  0.06
$$ \vspace{-0.07\textheight}$$
v_3 = 0.04
$$
\end{footnotesize}
\end{minipage}
\begin{minipage}[b]{0.18\textwidth} 
\begin{footnotesize}
$$
v_2  =  0.04
$$ \vspace{-0.07\textheight}$$ 
v_3  =  0.06
$$
\end{footnotesize}
\end{minipage}
\begin{minipage}[b]{0.18\textwidth} 
\begin{footnotesize}
$$
v_2 = 0.06
$$ \vspace{-0.07\textheight}$$
v_3  =  0.06
$$
\end{footnotesize}
\end{minipage}
}
\centerline{
\begin{minipage}[b]{0.16\textwidth}$\Psi_{23} = 0$\\
\vspace{0.05\textheight}\end{minipage}
\includegraphics[width=0.18\textwidth]{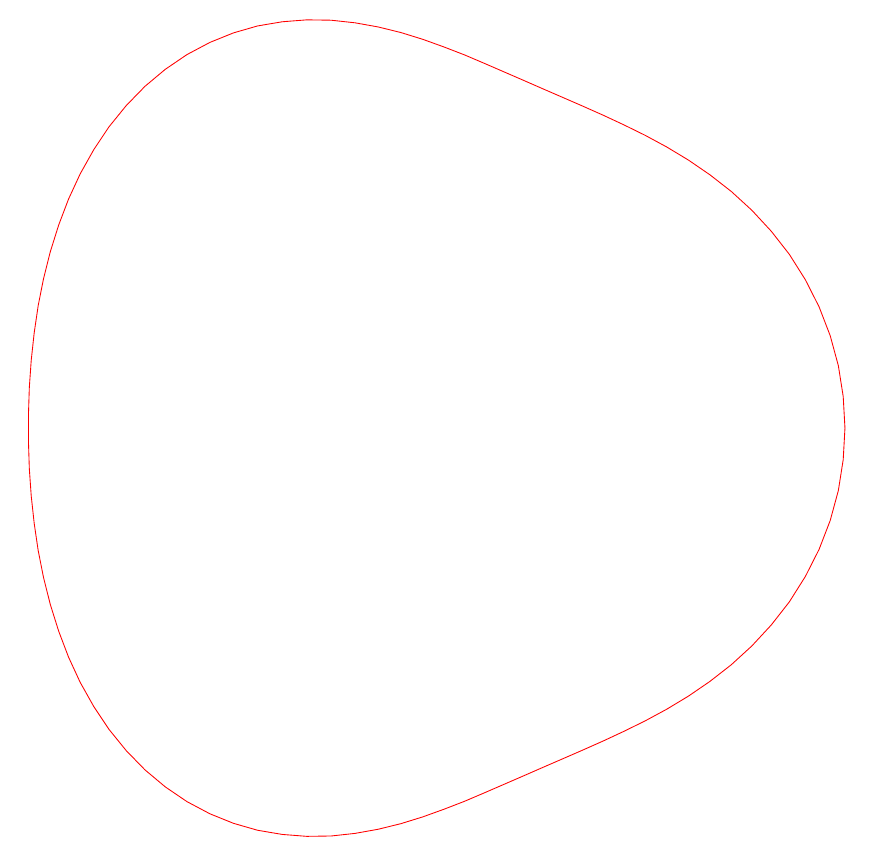}
\includegraphics[width=0.18\textwidth]{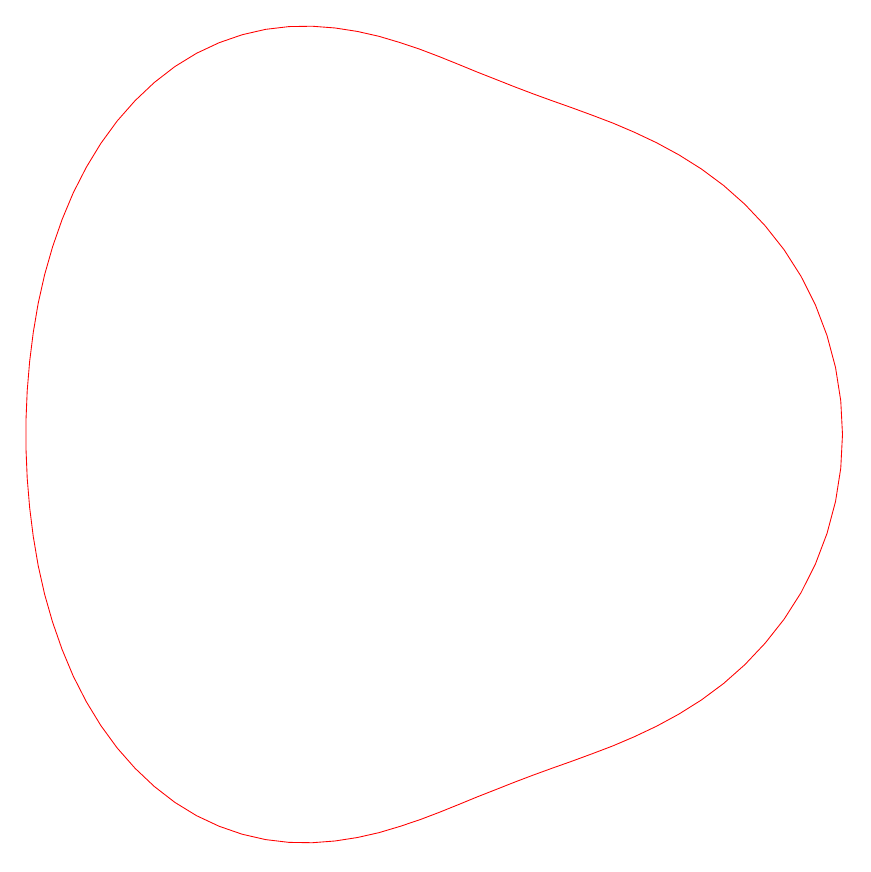}
\includegraphics[width=0.18\textwidth]{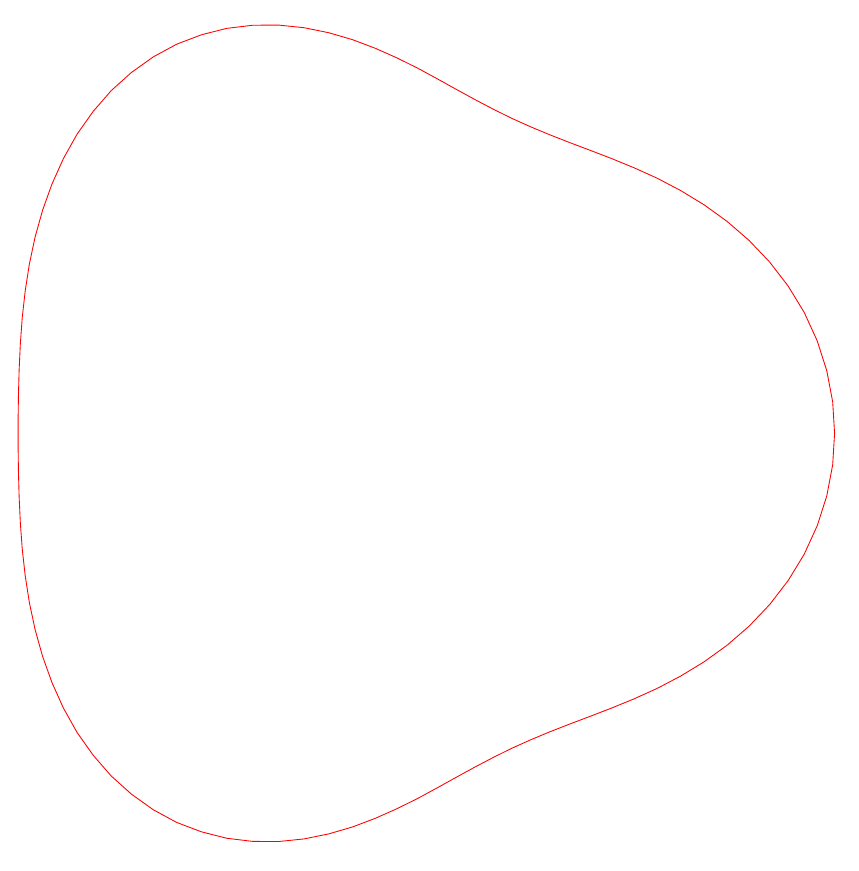}
\includegraphics[width=0.18\textwidth]{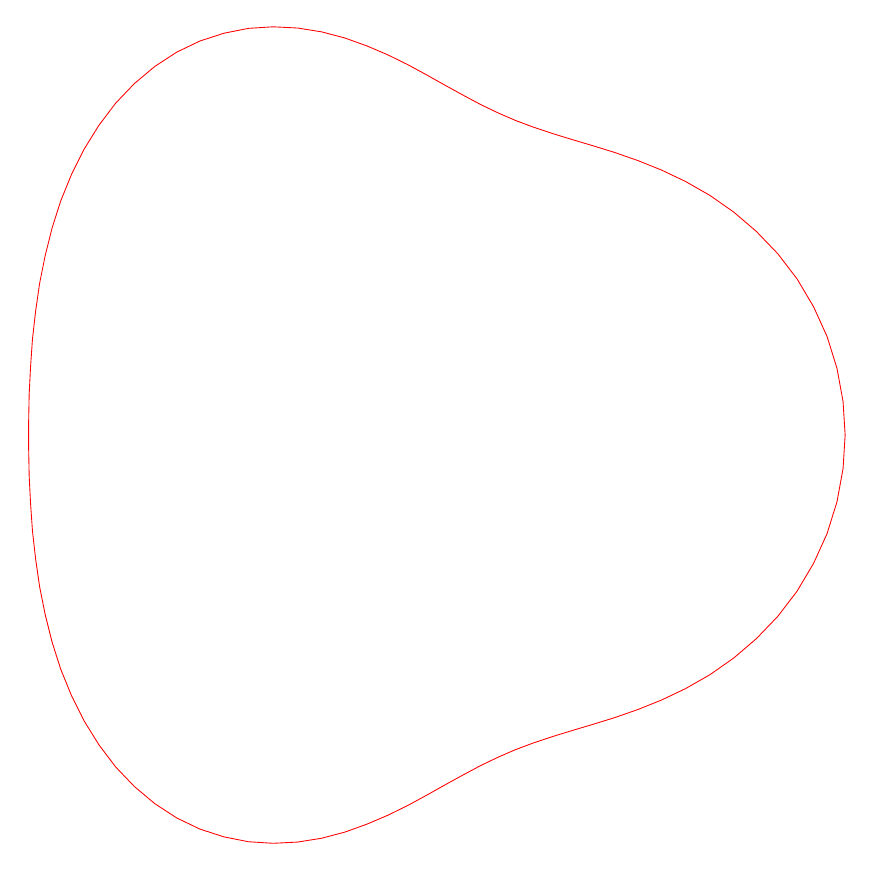}
}
\centerline{
\begin{minipage}[b]{0.16\textwidth}$\Psi_{23} = 0.7$\\
\vspace{0.05\textheight}\end{minipage}
\includegraphics[width=0.18\textwidth]{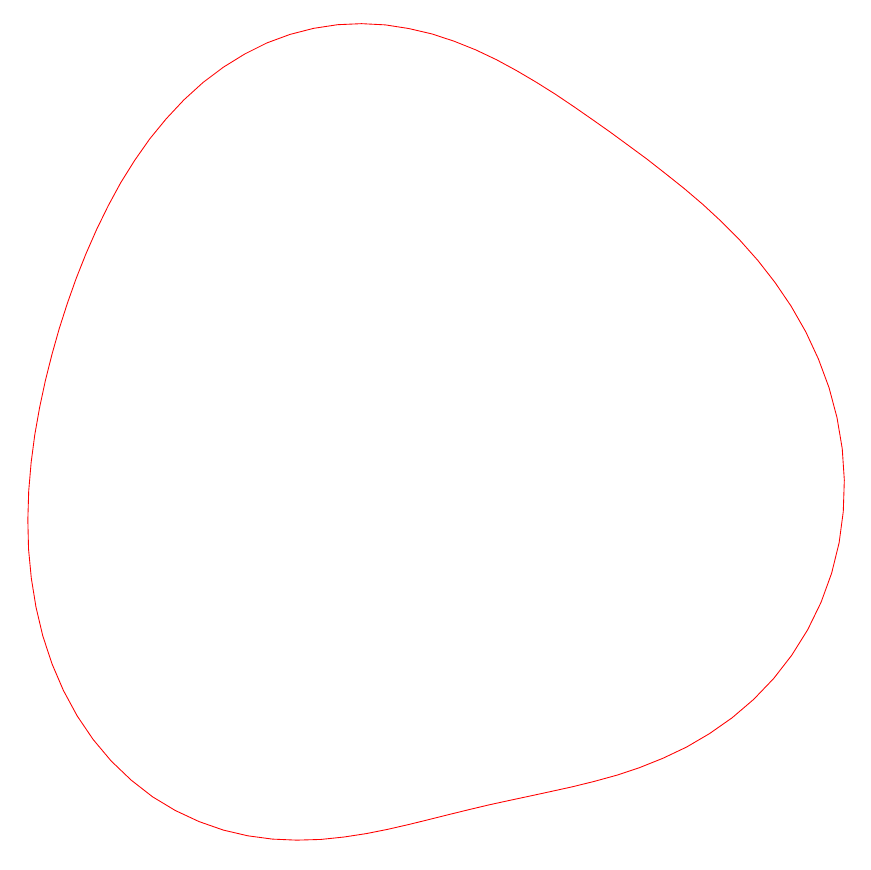}
\includegraphics[width=0.18\textwidth]{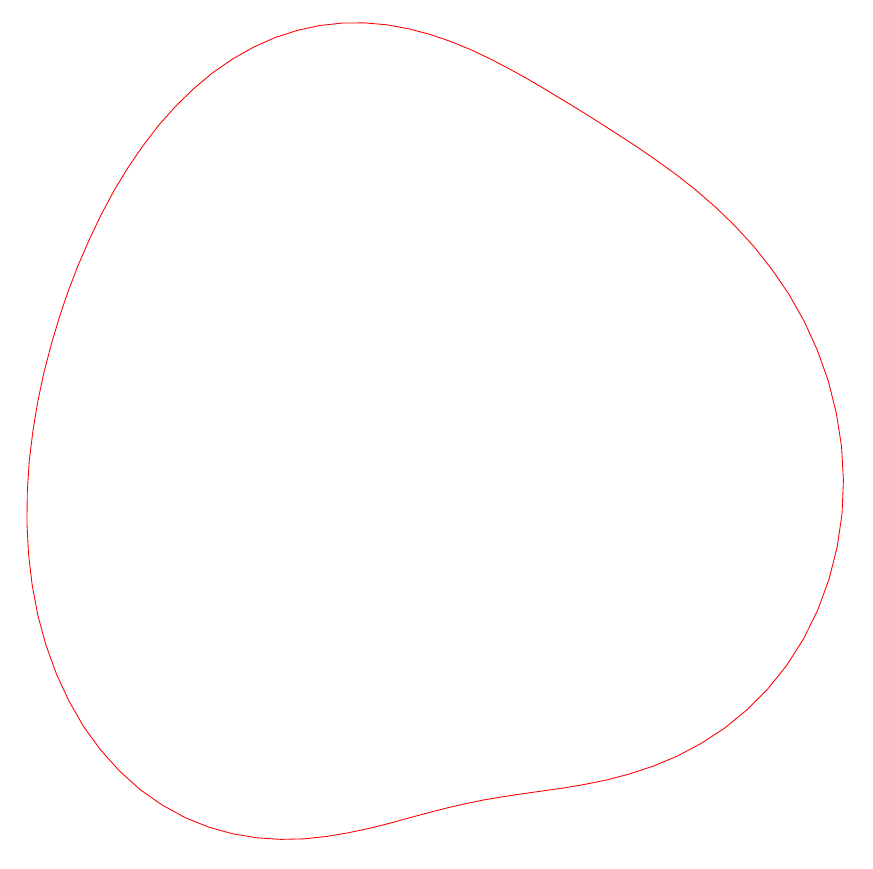}
\includegraphics[width=0.18\textwidth]{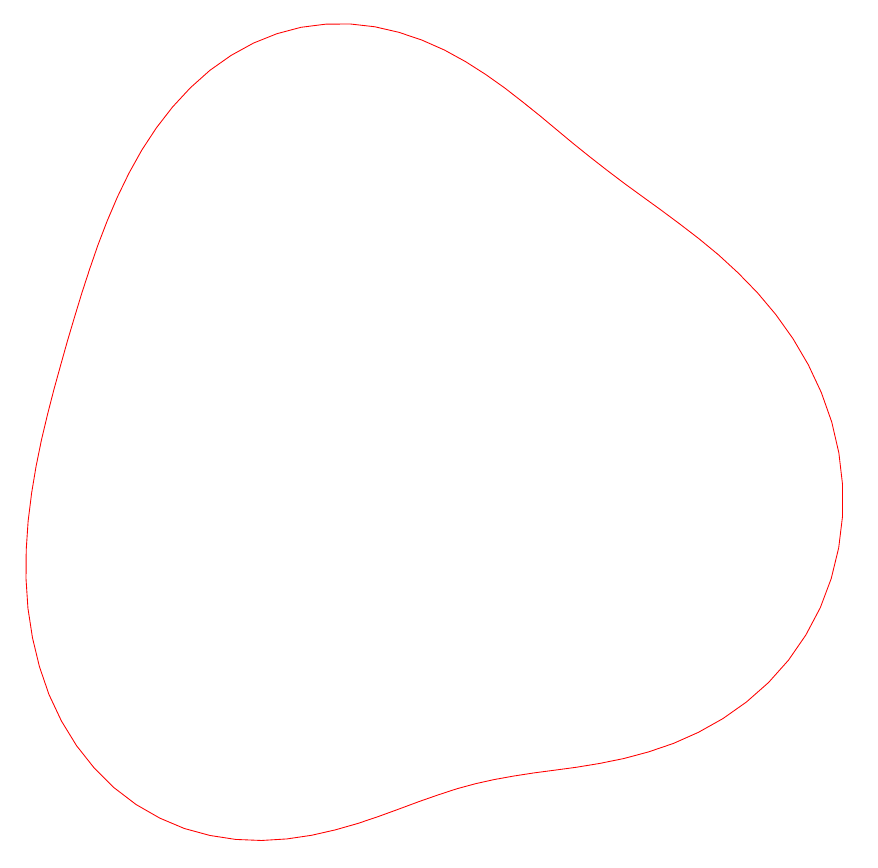}
\includegraphics[width=0.18\textwidth]{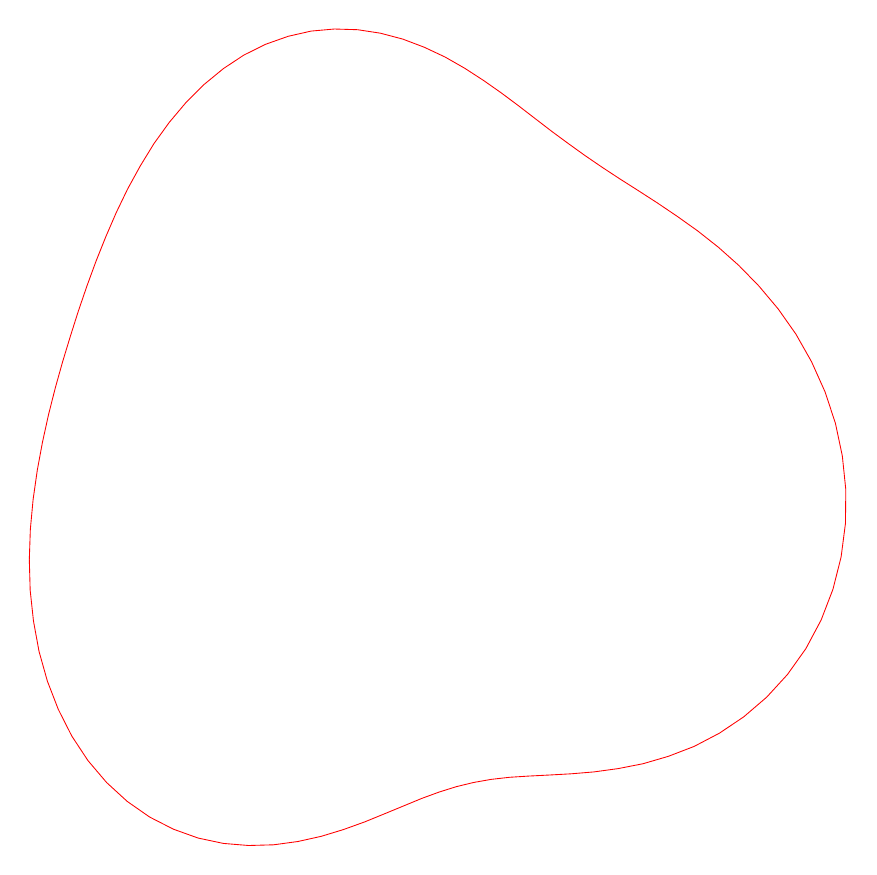}
}
\centerline{
\begin{minipage}[b]{0.16\textwidth}$\Psi_{23} = 1.57$\\
\vspace{0.05\textheight}\end{minipage}
\includegraphics[width=0.18\textwidth]{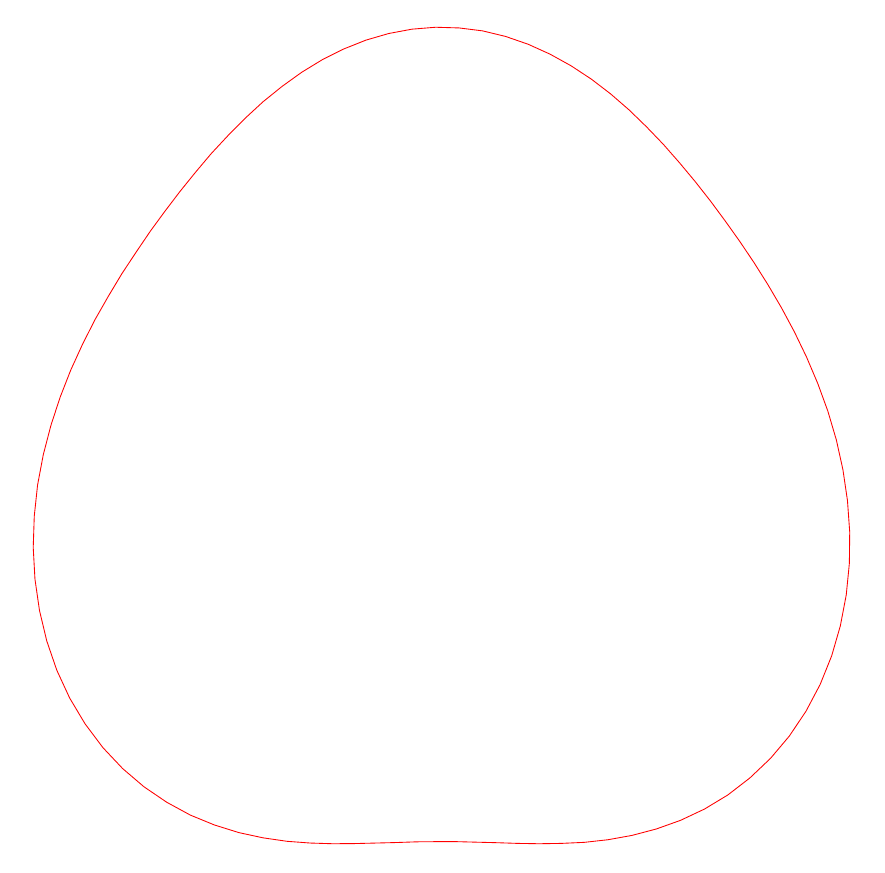}
\includegraphics[width=0.18\textwidth]{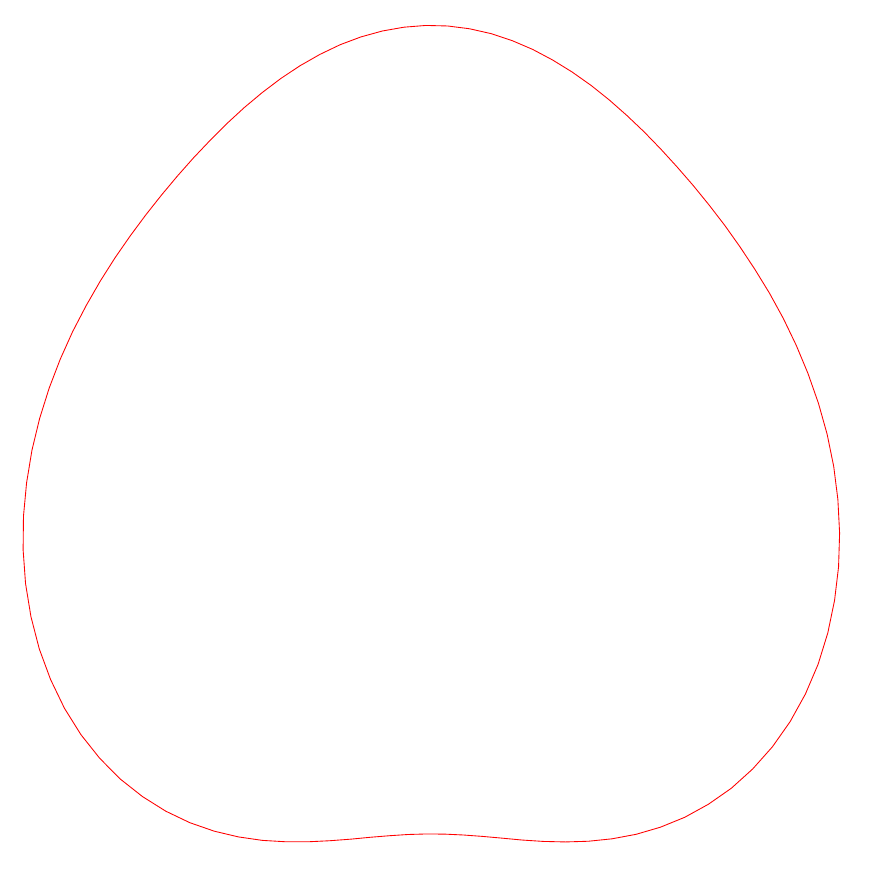}
\includegraphics[width=0.18\textwidth]{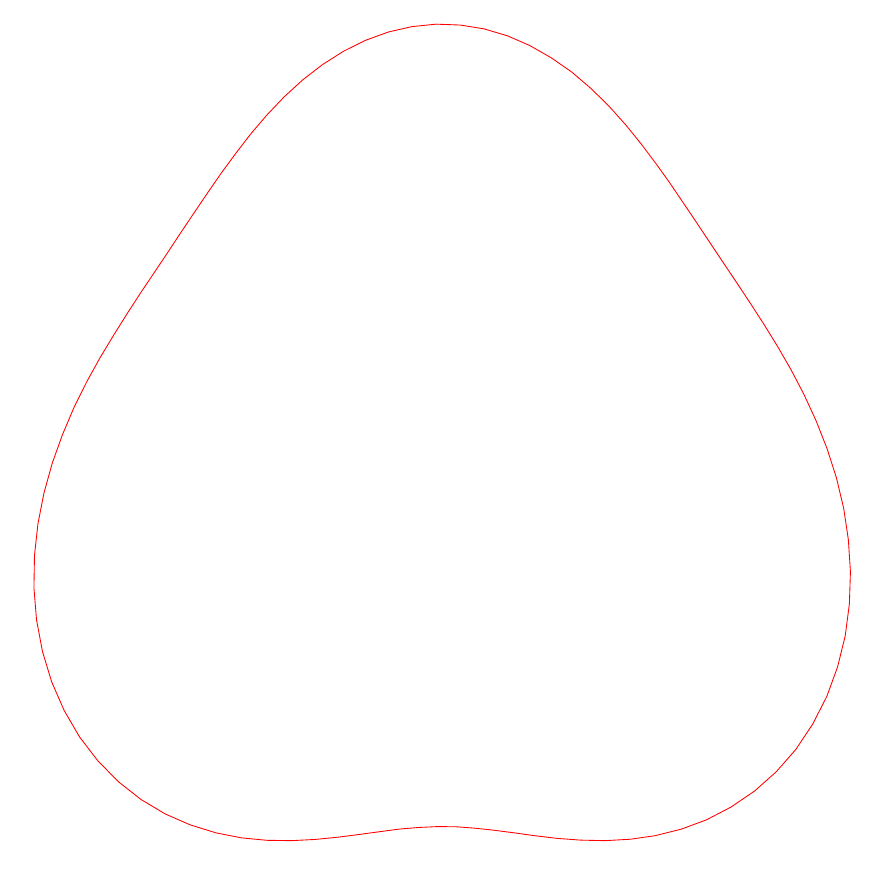}
\includegraphics[width=0.18\textwidth]{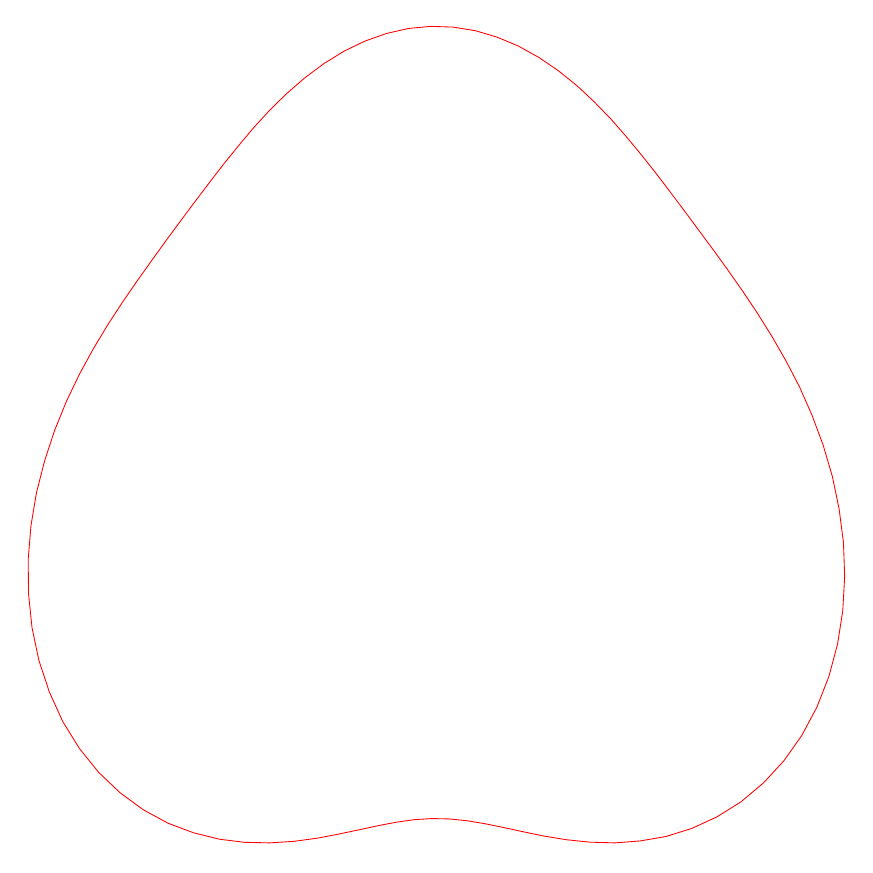}
}
\caption{Shapes with different azimuthal anisotropies drawn according to eq.~(\ref{e:anis}).
\label{f:shapes}
}
\end{figure}

A treatment similar in its spirit is known under the name Event Shape Engineering (ESE) \cite{Schukraft:2012ah}.
Let us explain the difference between ESE and ESS. In Fig.~\ref{f:shapes} we show twelve shapes for which the radius follows 
the prescription 
\begin{equation}
R(\phi) = R_0 \left (1 + 2v_2 \cos(2\phi) + 2v_3 \cos(3(\phi - \Psi_{23}))\right )\,  .
\label{e:anis}
\end{equation}
\begin{figure}
\centerline{\includegraphics[width=0.99\textwidth]{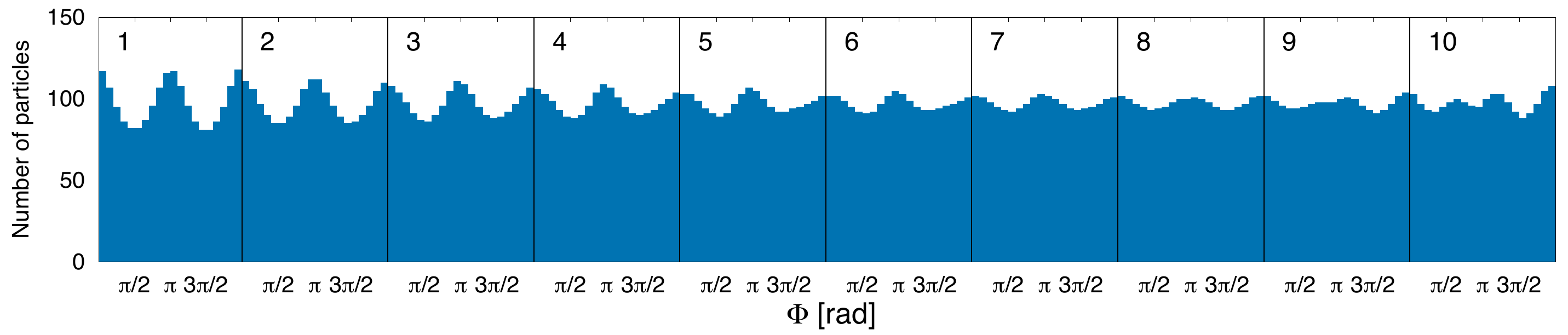}}
\caption{Histograms in azimuthal angle of ESS-sorted events generated by DRAGON.
\label{f:ESSsorted}
}
\end{figure}
In ESE one first selects the selection variable and then picks events with either highest or lowest values of that variable. Here, 
if e.g.~that variable was $v_2$ then in one group one would have shapes from the first  and the third column and in the other 
group the remaining shapes.  It is questionable if this is the natural way of grouping shapes. On the other hand, Fig.~\ref{f:ESSsorted}
shows average histograms of ESS-sorted events.
We clearly see that the sorting respects the richer structure of the events, including both second and third order oscillations.


\subsection{Femtoscopy studies with Event Shape Sorting}

\begin{figure}
\centerline{\includegraphics[width=0.46\textwidth]{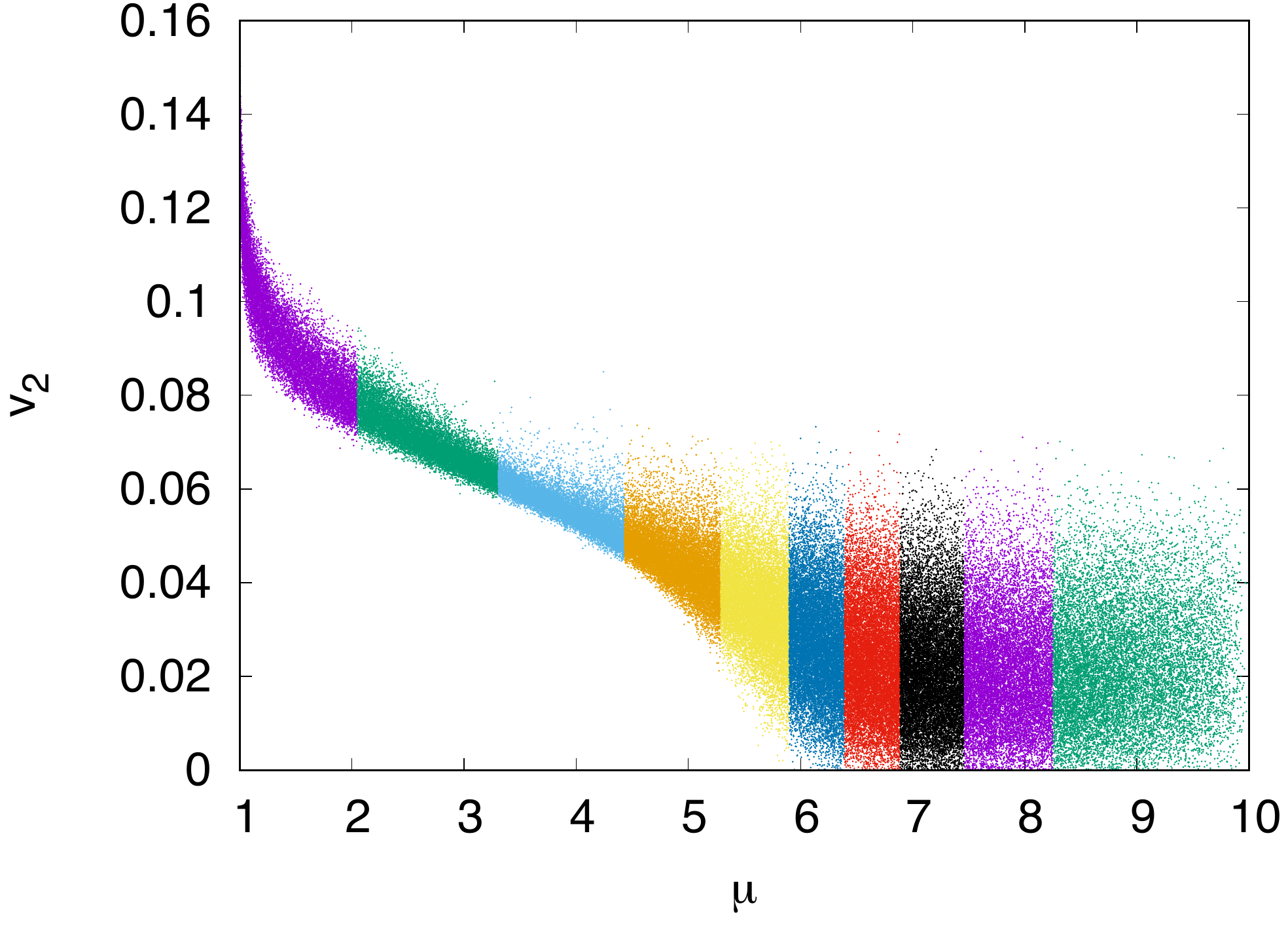}
\includegraphics[width=0.46\textwidth]{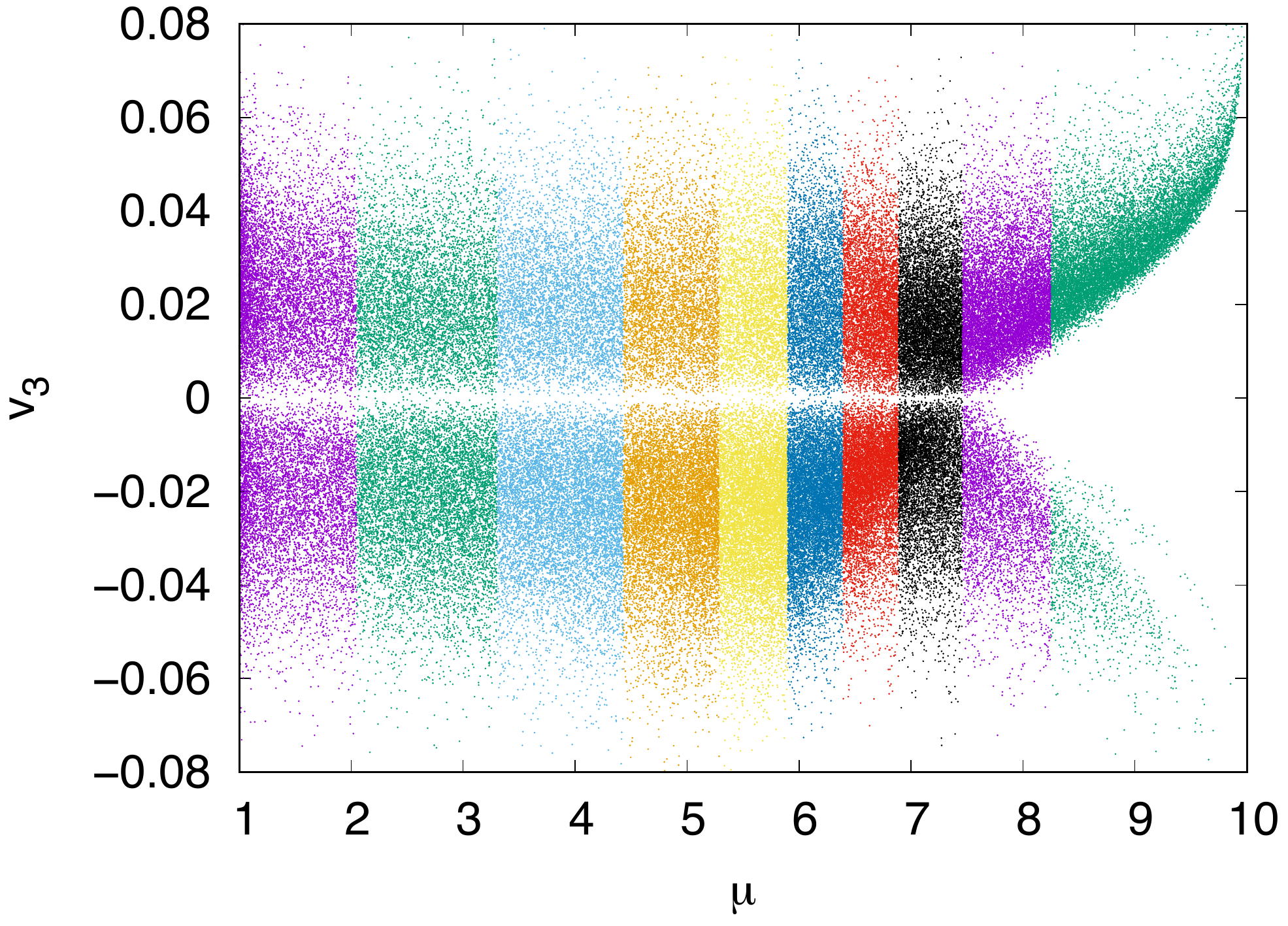}}
\caption{Coefficients $v_2$ (left) and $v_3$ (right) of individual events as depending on the sorting variable 
$\mu$. Different colours represent different event classes.
\label{f:flows}
}
\end{figure}

We have sorted 150~000 events generated by DRAGON \cite{Tomasik:2008fq}, which is a MC blast-wave generator 
with included resonances. It has been augmented to include azimuthal anisotropy according to eqs.~(\ref{e:Rani}) and 
(\ref{e:flowani}). The scatter plots in Fig.~\ref{f:flows} show the $v_2$ and $v_3$ as they depend on the sorting variable $\mu$
(see \cite{Kopecna:2015fwa}).  
The sorting goes dominantly according to $v_2$, but in classes 8--10 we observe an admixture of the third order anisotropy, as well. 

This picture is complemented by the azimuthal dependence of the correlation radii, which has been constructed for each event class 
separately. 
\begin{figure}
\centerline{\includegraphics[width=0.99\textwidth]{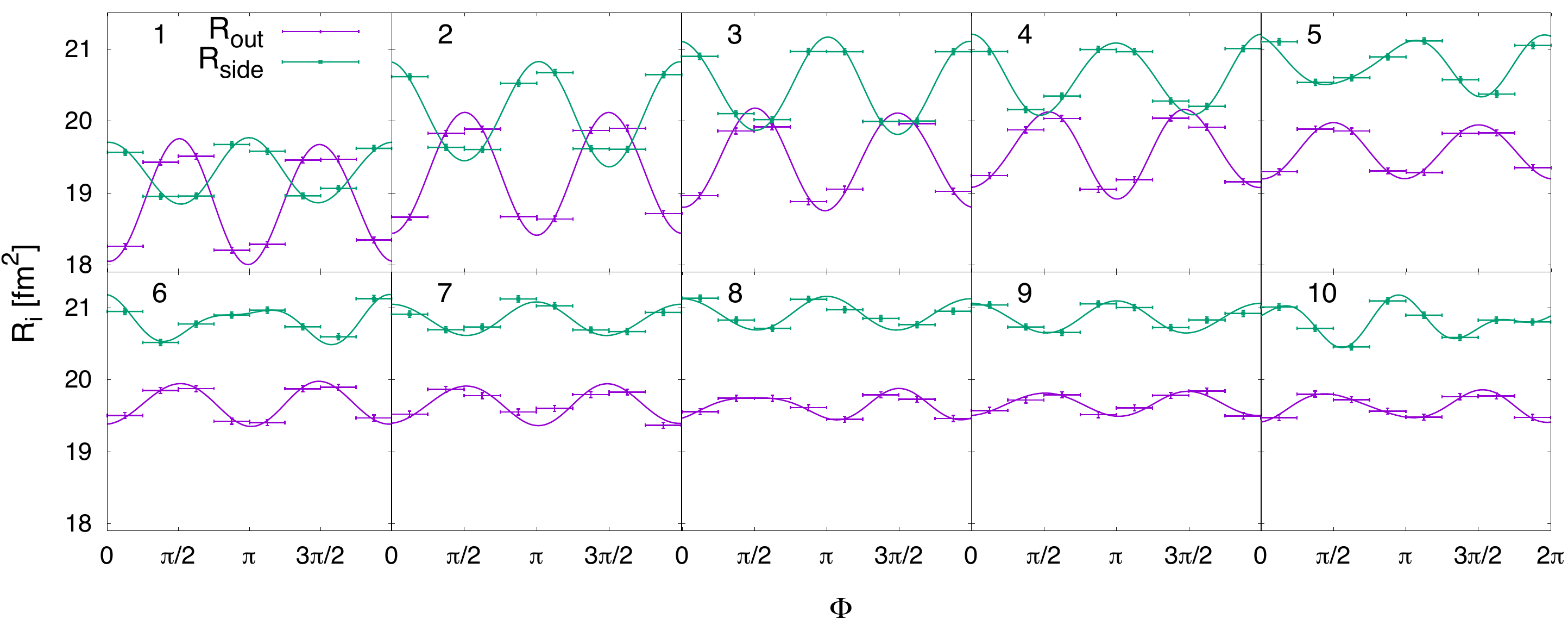}}
\caption{The azimuthal dependence of $R_{out}$ and $R_{side}$ from all event classes.
\label{f:Roscil}
}
\end{figure}
It is shown in Fig.~\ref{f:Roscil}. The third order component in the azimuthal dependence of the correlation radii 
is best seen in event classes 6, 9, and 10. 

Note that in experimental analyses so far one could only look at each oder of the oscillation separately. If all events were azimuthally rotated 
with respect to $n$-th order event plane ($n=2$ or 3) then only $n$-th order oscillation can be seen and the other order is averaged out.
The ESS technique thus offers the unique opportunity to observe all orders together at once.



\section{Conclusions}

Averaging over a large number of events has an impact on the shape of the correlation function. 
We showed that it can cast the correlation function into a shape of a  Levy-stable distribution.
It remains to be seen, to what extent the  present observations \cite{Kincses,balint} can be described by such averaging. 

The analysis could be made more exclusive and less influenced by averaging with the help of the 
Event Shape Sorting technique. We have demonstrated here, that the sorted classes differ also by the 
azimuthal dependence of the correlation radii. 

Note finally, that we have made the sorting algorithm \cite{Kopecna:2015fwa} available
\cite{nessie}.


\vspace*{-0.8em}
\subsubsection*{Acknowledgements}
\vspace*{-0.5em}
Supported by the grant 17-04505S of the Czech Science Foundation (GA\v{C}R).
BT also acknowledges support by VEGA 1/0348/18 (Slovakia).


\end{document}